\documentclass[showpacs,preprintnumbers,amsmath,amssymb]{revtex4}

\usepackage{graphicx}
\usepackage{dcolumn}
\usepackage{bm}

\begin{document}

\title{Blow-up Solutions to a Viscoelastic Fluid System and a Coupled
Navier-Stokes/Phase-Field System in $\mathbb{R}^2$\footnote{
Supported by National Science Foundation No.10971142.}}

\author{ZHAO Li-Yun £¨ÕÔÀøÔÅ£©$^1$\footnote{ E-mail: zhaoly69@gmail.com },
GUO Bo-Ling£¨¹ù°ØÁ飩$^2$\footnote{ E-mail: gbl@mail.iapcm.ac.cn},
 HUANG Hai-Yang£¨»Æº£Ñó£©$^1$\footnote{Corresponding author. E-mail: hhywsg@bnu.edu.cn(Huang H.Y.) }}

\affiliation{$^1$School of Mathematical Sciences and Key Laboratory
of Mathematics and Complex Systems (Ministry of Education), Beijing
Normal University, Beijing 100875 \\ $^2$Institute of Applied
Physics and Computational Mathematics, Nonlinear Center for Studies,
PO Box 8009, Beijing 100088}


\begin{abstract}
The explicit solutions to both the Oldroyd-B model with infinite
Weissenberg number and the coupled Navier-Stokes/phase-field system
are constructed by the method of separation of variables. It is
found that the solutions blow up in finite time.
\end{abstract}

\pacs{02.30.Gp,  02.30.Jr}
\maketitle

The Oldroyd-B model with infinite Weissenberg number for an
incompressible viscoelastic fluid system in $\mathbb{R}^n$ takes the
form (see \cite{LLZhang05,ZhengLZhou08}):{\small \begin{eqnarray}
\label{O-Bmodel}
{\bf u}_t+({\bf u}\cdot\nabla){\bf u}+\nabla p&=&\nu\Delta{\bf u}+\lambda\nabla\cdot(\mathcal{F}\mathcal{F}^{T}),\quad\quad\\
\nabla\cdot {\bf u}&=&0, \label{incompress1}\\
\mathcal{F}_t+{\bf u}\cdot\nabla \mathcal{F}&=&\nabla {\bf
u}\mathcal{F},\label{F} \end{eqnarray} }where ${\bf u}$ represents
the velocity vector in $\mathbb{R}^n$, $p$ is a scalar function
denoting the pressure and $\mathcal{F}$ is a $n\times n$ matrix
denoting the deformation tensor. $\nu$ and $\lambda$ are positive
constants, denoting the kinetic viscosity and the competition
between kinetic energy and elastic energy, respectively.

Let ${\bf x}$ be the Eulerian coordinates and ${\bf X}$ the
Lagrangian coordinates. The flow map ${\bf x}(X,t)$ is defined by
\begin{equation*}{\bf x}_t(X,t)={\bf u}({\bf x}(X,t),t),\quad {\bf x}(X,0)=X.
\end{equation*} The deformation tensor
$\tilde{\mathcal{F}}(X,t)=\frac{\partial{\bf x}}{\partial{\bf
X}}(X,t)$.
 In the Eulerian coordinates, the corresponding deformation tensor
$\mathcal{F}({\bf x},t)$ is defined as $\mathcal{F}({\bf
x}(X,t),t)=\tilde{\mathcal{F}}(X,t)$.
Using the chain rule, one obtains \eqref{F}, which stands for
{\small$\partial_t(\mathcal{F}_{ij})+
u_k\partial_{x_k}\mathcal{F}_{ij}=\partial_{x_k}u_i\mathcal{F}_{kj}$,
$1\leq i,j\leq n.$} Denote
{\small$(\nabla\cdot\mathcal{F})_j=\partial_{x_i}\mathcal{F}_{ij}$}.
Taking divergence of both sides of \eqref{F} and using
$\nabla\cdot{\bf u}=0$, one gets the transport equation for
$\nabla\cdot\mathcal{F}$ as
{\small\begin{equation}(\nabla\cdot\mathcal{F})_t+{\bf
u}\cdot\nabla(\nabla\cdot\mathcal{F})=0.\label{div-F}\end{equation}}
We assume naturally the initial datum $\mathcal{F}_0=I$, where $I$
is the identity matrix. Thus $\nabla\cdot\mathcal{F}_0=0$ and
$\det\mathcal{F}_0=1$. Then \eqref{div-F} yields{\small
\begin{equation} \label{zero divergence of deformation tensor F}
\nabla\cdot\mathcal{F}=0,\ \ \forall\, t \geq 0
,\end{equation}}which implies that in the 2-D case ($n=2$), there
exists a vector function $\bm\phi=(\phi_1,\phi_2)$ such that
{\small\begin{eqnarray}\mathcal{F}&=&\nabla^{\perp}\bm\phi=\left(
                              \begin{array}{c}
                                -\partial_{x_2} \bm\phi \\
                                 \partial_{x_1} \bm\phi \\
                              \end{array}
                            \right)  \nonumber\\
                            &=& \left(
                              \begin{array}{cc}
                                -\partial_{x_2} \phi_1 & -\partial_{x_2} \phi_2\\
                                 \partial_{x_1} \phi_1 & \partial_{x_1} \phi_2\\
                              \end{array}
                            \right).\label{F-2D}\end{eqnarray}}Therefore, \eqref{O-Bmodel}-\eqref{F} in $\mathbb{R}^2$ can be
transformed into an equivalent form as follows (see
\cite{LLZhang05}): {\small\begin{eqnarray} {\bf u}_t+({\bf
u}\cdot\nabla){\bf u}+\nabla P&=&\nu\Delta{\bf
u}-\lambda\nabla\cdot(\nabla\bm\phi\otimes\nabla\bm\phi),
\label{Navier Stokes in Oldroyd-B}\\ \nabla\cdot{\bf u}&=&0, \\
\bm\phi_t+({\bf u}\cdot\nabla)\bm\phi&=&0,\label{trans equ in
Oldroyd} \end{eqnarray}}where
$P=p-|\partial_{x_1}\bm\phi|^2-|\partial_{x_2}\bm\phi|^2$ and
$\nabla\bm\phi\otimes\nabla\bm\phi$ is a $2\times 2$ matrix whose
$(i,j)$-th entry is
$\partial_{x_i}\bm\phi\cdot\partial_{x_j}\bm\phi$ for $1\leq i,j\leq
2$.

Due to $\det\mathcal{F}_0=1$ and \eqref{F-2D}, we assume the initial
datum $\bm\phi_0$ of system \eqref{Navier Stokes in
Oldroyd-B}-\eqref{trans equ in Oldroyd} satisfies
$\det(\nabla^\perp\bm\phi_0)=1$. It was proved in \cite{LLZhang05}
that the system \eqref{Navier Stokes in Oldroyd-B}-\eqref{trans equ
in Oldroyd} has global classical solutions on the whole plane
$\mathbb{R}^2$, provided that the initial data $({\bf
u}_0,\bm\phi_0)$ satisfies the following assumptions:

(A1)  ${\bf u}_0$ is near zero; \quad

(A2) $\det(\nabla^\perp\bm\phi_0)=1$, $\bm\phi_0$ is close to the
vector ${\bf a}=(-x_2,x_1)$;

(A3) ${\bf u}_0\in H^k(\mathbb{R}^2)$ and
$\nabla^\perp(\bm\phi_0-{\bf
a})\in H^k(\mathbb{R}^2)$, $k\in\mathbb{N}$, $k\geq 2$.\\
Furthermore, the authors in \cite{ZhengLZhou08} proved that the
system \eqref{O-Bmodel}-\eqref{F} has global smooth solutions in the
whole space $\mathbb{R}^n$ ($n=2,3$) if $({\bf u}_0,\mathcal{F}_0)$
satisfies similar conditions to (A1)-(A3). Specifically, the three
assumptions are:

$(A1)'$ ${\bf u}_0$ is near zero;

$(A2)'$  $\det\mathcal{F}_0=1$, $\nabla\cdot\mathcal{F}_0=0$,
$\mathcal{F}_0$ is near the identity matrix;

$(A3)'$ ${\bf u}_0, \mathcal{F}_0-I\in H^k(\mathbb{R}^2)$,
$k\in\mathbb{N}$, $k\geq 2$.

\noindent There are no results for global existence if the magnitude
of ${\bf u}_0$ is large.

Due to the above result in \cite{LLZhang05}, initial data of a blow
up solution to \eqref{Navier Stokes in Oldroyd-B}-\eqref{trans equ
in Oldroyd} could not satisfy every condition of (A1)-(A3). In the
first part of this letter, we construct such a blow up solution of
\eqref{Navier Stokes in Oldroyd-B}-\eqref{trans equ in Oldroyd} that
its initial data satisfies (A2), but violates (A1) and (A3). We use
the method of separation of variables. Assume
\begin{eqnarray} &&u_1({\bf x},t)=x_1f(t),\, u_2({\bf x},t)=-x_2f(t), \quad\label{u-assume} \\
&&P({\bf x},t)=P_1({\bf x})g(t),\\
&&\phi_1({\bf x},t)=-x_2h_1(t)+x_1g_1(t),\\
&&\phi_2({\bf x},t)=x_1h_2(t)-x_2g_2(t). \label{phi-assume}
 \end{eqnarray}
Then the incompressibility condition $\nabla\cdot{\bf u}=0$ holds.
Since $\mathcal{F}_0=\nabla^\perp\bm\phi_0$ and $\mathcal{F}_0$ is
the identity matrix, we have \begin{eqnarray} \nabla^\perp\bm\phi_0=
\left.\left(
                              \begin{array}{cc}
                                -\partial_{x_2} \phi_1 & -\partial_{x_2} \phi_2\\
                                 \partial_{x_1} \phi_1 & \partial_{x_1} \phi_2\\
                              \end{array}
                            \right)\right|_{t=0}\nonumber\\
                        = \left.\left(
                              \begin{array}{cc}
                                h_1(t) & g_2(t)\\
                                g_1(t) & h_2(t)\\
                              \end{array}
                            \right)\right|_{t=0}
                            = \left(
                              \begin{array}{cc}
                               1 & 0\\
                               0 & 1\\
                              \end{array}
                            \right) .\label{ID} \end{eqnarray}
In addition, we have $\Delta{\bf u}=0$ and
$\nabla\cdot(\nabla\bm\phi\otimes\nabla\bm\phi)=0$.

Substituting \eqref{u-assume}-\eqref{phi-assume} into \eqref{Navier
Stokes in Oldroyd-B}, we get \begin{eqnarray*}
x_1f'(t)+x_1f^2(t)+\frac{\partial P_1({\bf x})}{\partial x_1}g(t)&=&0,\\
-x_2f'(t)+x_2f^2(t)+\frac{\partial P_1({\bf x})}{\partial
x_2}g(t)&=&0.\end{eqnarray*} By separation of variables, we obtain
\begin{eqnarray}
\frac{f'(t)+f^2(t)}{g(t)}&=&\frac{-1}{x_1}\frac{\partial
P_1({\bf x})}{\partial x_1}=-\alpha,\quad \label{f-1}\\
\frac{f'(t)-f^2(t)}{g(t)}&=&\frac{1}{x_2}\frac{\partial P_1({\bf
x})}{\partial x_2}=-\beta,\label{f-2} \end{eqnarray} where $\alpha$,
$\beta$ are constants. Obviously, $P_1({\bf x})$ has the form of
$P_1({\bf x})=\frac{1}{2}(\alpha x_1^2-\beta x_2^2)$ up to a
constant. Suppose $f(0)=f_0$. Solving Eqs. \eqref{f-1} and
\eqref{f-2}, we obtain
\begin{eqnarray}
f(t)&=&\frac{f_0}{1-\frac{\alpha+\beta}{\alpha-\beta}f_0t},\label{f-soln}\\
g(t)&=&\frac{2}{\beta-\alpha}f^2(t)\nonumber\\
&=&\frac{2f_0^2}{(\beta-\alpha)\left(1-\frac{\alpha+\beta}{\alpha-\beta}f_0t\right)^2}.\label{g}
\end{eqnarray}

Substituting \eqref{u-assume}-\eqref{phi-assume} into \eqref{trans
equ in Oldroyd}, we obtain \begin{eqnarray*}
x_1(g'_1(t)+f(t)g_1(t))-x_2(h'_1(t)-f(t)h_1(t))=0,\\
x_1(h'_2(t)+f(t)h_2(t))-x_2(g'_2(t)-f(t)g_2(t))=0,\end{eqnarray*}
for arbitrary $x_1$ and $x_2$. Therefore, \begin{eqnarray*}
\, g'_1(t)+f(t)g_1(t)=0,\quad h'_1(t)-f(t)h_1(t)=0,\\
\,   h'_2(t)+f(t)h_2(t)=0,\quad g'_2(t)-f(t)g_2(t)=0,
\end{eqnarray*} which are subject to the initial condition
\eqref{ID}. Hence, we get
\begin{eqnarray} g_1(t)&=& g_1(0)e^{-\int^t_0f(s)ds}\equiv 0,\nonumber\\
g_2(t)&=& g_2(0)e^{\int^t_0f(s)ds}\equiv 0, \nonumber\\
h_1(t)&=& h_1(0)e^{\int^t_0f(s)ds}=e^{\int^t_0f(s)ds}\nonumber\\
&=&
\left|1-\frac{\alpha+\beta}{\alpha-\beta}f_0t\right|^{\frac{\beta-\alpha}{\alpha+\beta}},\label{h-1}\\
h_2(t)&=& h_2(0)e^{-\int^t_0f(s)ds}=e^{-\int^t_0f(s)ds},\nonumber\\
&=&\left|1-\frac{\alpha+\beta}{\alpha-\beta}f_0t\right|^{\frac{\alpha-\beta}{\alpha+\beta}}.\label{h-2}
\end{eqnarray}

Finally, substituting \eqref{f-soln}, \eqref{g}, \eqref{h-1} and
\eqref{h-2} into \eqref{u-assume}-\eqref{phi-assume}, we find the
following explicit solutions of \eqref{Navier Stokes in
Oldroyd-B}-\eqref{trans equ in Oldroyd} with initial datum
$\bm\phi_0$ satisfying (A2):
{\small\begin{equation}\label{soln1}\left\{\begin{array}{l} {\bf
u}({\bf x},t)
     =\left(\begin{array}{c}
      \frac{x_1f_0}{1-\frac{\alpha+\beta}{\alpha-\beta}f_0t}  \\
      \frac{-x_2f_0}{1-\frac{\alpha+\beta}{\alpha-\beta}f_0t}
        \end{array} \right),\\
P({\bf x},t)=\frac{\left(\alpha x_1^2-\beta x_2^2\right)f_0^2}
{(\beta-\alpha)\left(1-\frac{\alpha+\beta}{\alpha-\beta}f_0t\right)^2},
\\\bm\phi({\bf x},t)=\left(\begin{array}{c}
      -x_2\left|1-\frac{\alpha+\beta}{\alpha-\beta}f_0t\right|^{\frac{\beta-\alpha}{\alpha+\beta}}        \\
      \quad x_1\left|1-\frac{\alpha+\beta}{\alpha-\beta}f_0t\right|^{\frac{\alpha-\beta}{\alpha+\beta}}   \\
         \end{array} \right),
\end{array} \right.
 \end{equation}}
where $f_0$, $\alpha$ and $\beta$ are constants.

With the use of \eqref{F-2D} and \eqref{soln1}, the corresponding
deformation tensor $\mathcal{F}$ becomes \begin{eqnarray*}
\mathcal{F}&=&\mathcal{F}(t)\\
&=&\left(\begin{array}{cc}
\left|1-\frac{\alpha+\beta}{\alpha-\beta}f_0t\right|^{\frac{\beta-\alpha}{\alpha+\beta}}&0 \\
0&\left|1-\frac{\alpha+\beta}{\alpha-\beta}f_0t\right|^{\frac{\alpha-\beta}{\alpha+\beta}}
\end{array} \right).\end{eqnarray*}

If $\frac{\alpha+\beta}{\alpha-\beta}f_0>0$, $\alpha+\beta\neq 0$
and $\alpha-\beta\neq 0$, \eqref{soln1} will blow up at time
$t^*=\frac{\alpha-\beta}{(\alpha+\beta)f_0}$. Recall the conditions
(A1)-(A3) which guarantee the global existence of smooth solutions,
here (A1) and (A3) are violated in \eqref{soln1}. As $t\rightarrow
t^*$, $1-\frac{\alpha+\beta}{\alpha-\beta}f_0t\rightarrow 0$, one
diagonal element of $\mathcal{F}$ tends to zero, while the other
tends to infinity. This means the viscoelastic fluid is squeezed in
one spatial direction, but stretched in the other direction.

Below we will construct some blow up solutions to the following
coupled Navier-Stokes/Allen-Cahn system :
{\small\begin{eqnarray}{\bf u}_t+({\bf u}\cdot\nabla){\bf u}+\nabla
P&=&\nu\Delta{\bf
u}-\lambda\nabla\cdot(\nabla\phi\otimes\nabla\phi),\label{u-eqn}\\
\nabla\cdot{\bf u}&=&0,  \label{incompress}\\
\phi_t+({\bf
u}\cdot\nabla)\phi&=&\gamma(\Delta\phi-f(\phi)).\label{Allen-cahn}\end{eqnarray}}
\eqref{Allen-cahn} is the  advective Allen-Cahn equation. If it is
replaced with the advective Cahn-Hilliard equation:
{\small\begin{equation} \phi_t+({\bf u}\cdot\nabla)\phi =
-\gamma\Delta(\Delta\phi-f(\phi)), \label{Cahn-H}
\end{equation}}then the system \eqref{u-eqn}, \eqref{incompress},
\eqref{Cahn-H} is a coupled Navier-Stokes/Cahn-Hilliard system. If
$\gamma=0$ and the scalar function $\phi$ in
\eqref{Allen-cahn}/\eqref{Cahn-H} is taken as a vector function
$\bm\phi=(\phi_1,\phi_2)$ , then we formally arrive at \eqref{Navier
Stokes in Oldroyd-B}-\eqref{trans equ in Oldroyd}.

The system \eqref{u-eqn}-\eqref{Allen-cahn}/\eqref{Cahn-H} are two
types of  Navier-Stokes/phase-field model, describing the motion of
incompressible viscous two-phase fluids (see, for example,
\cite{JCP1999,LS03,ZWH2009}). The two fluids are separated by a thin
interface of width $\varepsilon>0$, which is a small constant. ${\bf
u}$ represents the velocity field of the mixture, $P$ is the
pressure, and $\phi$ is the phase function, taking the value 1 in
one bulk phase and -1 in the other. In the interfacial region,
$\phi$ varies rapidly and smoothly. $\nabla\phi\otimes\nabla\phi$
denotes the induced elastic stress, which is a $2\times 2$ matrix
whose $(i,j)$-th entry is $\frac{\partial\phi}{\partial
x_i}\frac{\partial\phi}{\partial x_j}$ for $1\leq i,j\leq 2$.
$f(\phi)=\frac{1}{\varepsilon^{2}}(\phi^3-\phi)$. $\nu$, $\lambda$
and $\gamma$ are positive constants, which denote the kinetic
viscosity constant, the mixing energy density and the mobility,
respectively.

We want to find blow up solutions of
\eqref{u-eqn}-\eqref{Allen-cahn}/\eqref{Cahn-H} in $\mathbb{R}^2$,
which are based on solutions of incompressible Navier-Stokes
equations: \begin{equation}\label{N-S} \left\{\begin{array}{l} {\bf
u}_t+({\bf u}\cdot\nabla){\bf u}+\nabla
p= \nu\Delta{\bf u},\\
\nabla\cdot{\bf u}=0.
\end{array} \right.
\end{equation}

An explicit blow up solution of \eqref{N-S} in $\mathbb{R}^2$ is
constructed as follows (see \cite{Guo-CPL}): {\small
\begin{equation}\label{N-S-sn} \left\{\begin{array}{ll}
u_1({\bf x},t)=&\frac{1}{\sqrt{T-t}}\left(-1+c_1\exp\left(\frac{s^2}{8\nu(T-t)}\right.\right.\\
               &\left.\left.-\frac{s}{\nu\sqrt{T-t}}+c_3\right)\right),\\
u_2({\bf x},t)=&\frac{1}{\sqrt{T-t}}\left(-1-c_1\exp\left(\frac{s^2}{8\nu(T-t)}\right.\right.\\
               &\left.\left.-\frac{s}{\nu\sqrt{T-t}}+c_3\right)\right),\\
p({\bf x},t)=&\frac{s}{2(T-t)^{3/2}}+\frac{c_2}{T-t},\quad\quad\quad
\end{array} \right.\end{equation}}where $s=x_1+x_2$, $T$ is a positive constant
and $c_i \,(i=1,2,3)$ are constants.

We suppose \begin{equation}\phi({\bf
x},t)=\Phi(s+f(t)),\label{phi}\end{equation} where $\Phi(\cdot)$ and
$f(\cdot)$ are non-constant $C^1$ functions to be determined. Then
$\partial_{x_1}\phi=\partial_{x_2}\phi=\Phi'$, and
$\nabla\cdot(\nabla\phi\times\nabla\phi)$ can be rewritten as
$2\nabla(\Phi')^2$. Hence, comparing \eqref{N-S} with \eqref{u-eqn},
we observe that $u_1$, $u_2$ in \eqref{N-S-sn} and
$P=p-2\lambda(\Phi')^2$ satisfy \eqref{u-eqn}.

Next we consider \eqref{Allen-cahn} in a simple case $\gamma=0$.
Substituting \eqref{phi} into
\begin{equation}\label{pure-transport}\phi_t+({\bf
u}\cdot\nabla)\phi=0,\end{equation} and using $\Phi'\neq 0$, we get
\begin{equation} f'(t)-\frac{2}{\sqrt{T-t}}=0. \label{f-ode}\end{equation}
Solving \eqref{f-ode}, we have $f(t)=-4\sqrt{T-t}+C$, where $C$ is a
constant. Therefore, $\phi({\bf x},t)=\Phi(s-4\sqrt{T-t}+C)$ is a
solution to \eqref{pure-transport}.

When $\gamma>0$, note that $\tanh(\frac{x_1+x_2}{2\varepsilon})$ is
a solution to $\Delta\phi-f(\phi)=0$. Hence, taking
$\Phi(\cdot)=\tanh(\frac{\cdot}{2\varepsilon})$, namely, $\phi({\bf
x},t)=\tanh(\frac{x_1+x_2-4\sqrt{T-t}+C}{2\varepsilon})$, we get a
solution to \eqref{Allen-cahn}/\eqref{Cahn-H}.

Finally, we obtain a blow up solution of
\eqref{u-eqn}-\eqref{Allen-cahn}/\eqref{Cahn-H} in $\mathbb{R}^2$:
\begin{equation*}\label{NSPF-sn} \left\{\begin{array}{ll}
u_1({\bf x},t)=&\frac{1}{\sqrt{T-t}}\left(-1+c_1\exp\left(\frac{s^2}{8\nu(T-t)}\right.\right.\\
               &\left.\left.-\frac{s}{\nu\sqrt{T-t}}+c_3\right)\right),\\
u_2({\bf x},t)=&\frac{1}{\sqrt{T-t}}\left(-1-c_1\exp\left(\frac{s^2}{8\nu(T-t)}\right.\right.\\
               &\left.\left.-\frac{s}{\nu\sqrt{T-t}}+c_3\right)\right),\\
\phi({\bf x},t)=&\tanh(\frac{s-4\sqrt{T-t}+c_4}{2\varepsilon}), \\
P({\bf x},t)=&\frac{s}{2(T-t)^{3/2}}-\frac{\lambda}{2\varepsilon^2}\cosh^{-4}(\frac{s-4\sqrt{T-t}}{2\varepsilon})\\
&+\frac{c_2}{T-t}.
\end{array} \right. \end{equation*}
The above solution can be easily generalized to the 3-D case:
\begin{equation*}\label{NSPF-sn} \left\{\begin{array}{ll}
u_1({\bf x},t)=&\frac{1}{\sqrt{T-t}}\left(-1+c_1\exp\left(\frac{s^2}{12\nu(T-t)}\right.\right.\\
               &\left.\left.-\frac{s}{\nu\sqrt{T-t}}+c_4\right)\right),\\
u_2({\bf x},t)=&\frac{1}{\sqrt{T-t}}\left(-1+c_2\exp\left(\frac{s^2}{12\nu(T-t)}\right.\right.\\
               &\left.\left.-\frac{s}{\nu\sqrt{T-t}}+c_4\right)\right),\\
u_3({\bf x},t)=&\frac{1}{\sqrt{T-t}}\left(-1-(c_1+c_2)\exp\left(\frac{s^2}{12\nu(T-t)}\right.\right.\\
               &\left.\left.-\frac{s}{\nu\sqrt{T-t}}+c_4\right)\right),\\
\phi({\bf x},t)=&\tanh(\frac{s-6\sqrt{T-t}+c_5}{\sqrt{6}\varepsilon}), \\
P({\bf x},t)=&\frac{s}{2(T-t)^{3/2}}-\frac{\lambda}{2\varepsilon^2}\cosh^{-4}(\frac{s-6\sqrt{T-t}}{\sqrt{6}\varepsilon})\\
&+\frac{c_3}{T-t}.
\end{array} \right. \end{equation*}


\medskip

\begin{thebibliography}{99}
\itemsep=0pt
\bibitem{LLZhang05} Lin F H, Liu C and Zhang P 2005 {\em Comm. Pure Appl. Math.} {\bf 58} 1

\bibitem{ZhengLZhou08} Lei Z, Liu C and Zhou Y 2008 {\em  Arch. Rational. Mech.
Anal.} {\bf 188} 371

\bibitem{JCP1999} Jacqmin D 1999 {\em J. Comput. Phys.} {\bf 155} 96

\bibitem{LS03} Liu C and Shen J 2003 {\em Physica D} {\bf 179} 211

\bibitem{ZWH2009} Zhao L Y, Wu H and Huang H Y 2009 {\em Commun. Math. Sci.} {\bf 7} 939

\bibitem{Guo-CPL} Guo B L, Yang G S and Pu X K 2008 {\em Chin. Phys. Lett.} {\bf 25} 2115

\end{thebibliography}
\end{document}